# Focusing of light beyond the diffraction limit

K. R. Chen*

*Department of Physics and Institute of Electro-optical Science and Engineering, National Cheng Kung University, 1 University Road, Tainan 70101, Taiwan, ROC.*

## Abstract

Diffraction limits the behaviour of light in optical systems and sets the smallest achievable line width at half the wavelength. With a novel subwavelength plasmonic lens to reduce the diffraction via an asymmetry and to generate and squeeze the wave functions, an incident light is focused by the aperture to a single-line with its width beyond the limit outside the near zone. The fields focused are radiative and capable of propagating to the far zone. The light focusing process, besides being of academic interest, is expected to open up a wide range of application possibilities.

Diffraction, as a general wave phenomenon which occurs whenever a traveling wave front encounters and propagates past an obstruction, was first referenced in the work of Leonardo da Vinci in the 1400s [1] and has being accurately described since Francesco Grimaldi in the 1600s [1]. Explanation based on a wave theory was not available until the 1800s [1]. The diffraction limit was the inspiration for Heisenberg's quantum uncertainty principle [2,3] that is a foundation of modern science; in fact, they can be deduced from each other [2-4]. The diffraction limit sets the smallest achievable line width at half the wavelength [1-4], which is the ultimate manipulability and resolution [1,4] of numerous diagnostic and fabrication instruments. The line width in a two dimensional system is $\lambda/2NA$, where $\lambda$ is the wavelength in vacuum, $NA = n\sin\theta$ is the numerical aperture, $n$ is the refractive index of the medium where the focused light locates, and $\theta$ is the convergence angle of the light.

The diffraction limit concerns travelling light that can propagate freely in free space, in contrast to the evanescent near-field [5,6] that needs a preferred plane or surface for propagation and cannot propagate freely in free space, such as occurs in a super lens [7-14]. The super lens was proposed [7] to reconstruct an electrostatic source distributed within a sub-limit size. It may bypass the diffraction limit for some particular cases of imaging and, as a result, has generated great attention [8-14]. Indeed, this is an interesting concept. But, what is considered and recovered within 0.1 $\lambda$ is the evanescent electrostatic near-field. Mathematically, the divergence of the electrostatic field and the Laplacian of the scalar potential are determined by the charge density according to Gauss's law and Poisson's equation, respectively, while the fields of propagating light concerned with the diffraction limit are decided by the spatial curl and temporal derivative equations (i.e., Faraday's and modified Ampere's equations) involving the scale of wavelength. Thus, a new understanding and a different approach are needed to surpass the fundamental physics limit and to produce a beyond-the-limit focusing of travelling light.

Both the theory of the diffraction limit of light and the Heisenberg's quantum uncertainty principle consider wave function expanded into reciprocal space and a situation where the scale of the eigenfunction is half the wavelength or larger. For quantum mechanics, the surface integral involving the wave function is assumed to vanish when taken over a very large surface or an infinite potential well. For the latter, the spatial eigenvalue (i.e., the wave number) is $k = m\pi/a$, where $m$ is a natural number (i.e., non-zero) and $a$ is the well width. For the ground state, the well width is half a wavelength, while the wave function is the corresponding sinusoidal function.

Conventional electromagnetic wave theory [15] indicates that the light cannot be transmitted through a subwavelength hole. However, the excitations of surface plasmon [16-18] on metallic surfaces and surface-plasmon-like modes [19-20] are claimed to enhance [21-22] the transmission of light and to beam [23] it. In fact, the light and the surface plasma are coupled and hence self-consistent within the slit. The wave function across the slit is close to a constant and drops sharply on the surface. This kind of function with $k = 0$ mode bounded within a sub-limit scale is not considered in the conventional theories and thus is not within their scope.

The innovative approach and physical mechanisms of the focusing aperture beyond the diffraction limit (FAB) of half the wavelength [24] are demonstrated here with the FAB lens including a metallic film with a double-slit and a patterned exit structure, as shown in Fig. 1(a). The width of each slit is smaller than half the wavelength, and thus the limit. For the purpose of providing a larger profile width of near field for excluding its involvement, the width of the central metal strip and the exit will be larger than the limit. Besides the generation of sub-limit wave function at the central area resulted from a polarized field conversion and the surface current, the transmitted light of sub-limit scale will be shown to be bent toward the center and focused to a single-line with its width well below the diffraction limit of half the wavelength.

Finite-Difference-Time-Domain (FDTD) simulation [25] is employed to verify the approach. A structured thin silver film with 20 and 2 grooves at the incident and exit sides, respectively, as shown in Fig. 1(b) is employed as our FAB lens. A simplified structure on silica substrate has been employed in an optical experiment [26] that verifies this approach, in addition to another experimental confirmation in the microwave range [27]. The refractive index of silver [28] used for $\lambda = 633$ nm is 0.134 + $i$ 3.99. The system has $1600 \times 600$ cells of the Yee space lattice with a unit cell length of 5 nm. The top of the silver film is at the $y = 400$ cell. The time, $t$, is normalized to the light period, and the time step is 0.005.

For better understanding of the approach and the physics involved, let us consider the dynamics of the focusing. The light with polarized electric $E_x$ and magnetic $H_z$ fields propagating downward is transmitted through the double-slit as enhanced by surface plasma. Then, due to the dynamics of the surface plasma on the non-symmetric metal at the slit exit, they are bent toward the central area in order to reduce the diffraction in the outward direction so as to preserve the sub-limit wave function, as indicated by the time-averaged $E_x$ field energy contour plot shown in Fig. 2(a). Besides being transmitted, the $H_z$ field of sub-limit scale can be produced by the generator at the central area due to the surface current and a subwavelength field conversion. The $E_y$ field, the polarized surface charge and the $x$ component of time-averaged Poynting vector, shown on Fig. 2(b), of the diffracted light from one slit are cancelled with those from the other. The cancellation converts the energy of the polarized surface charge and the $E_y$ field to increase the focused $H_z$ field [Fig. 2(c), at the time (and the phase) defined as $t = t_f$] and hence explains the behaviour of the $E_x$ field (Fig. 2a) at the central region so as to generate a sub-limit wave function there. The overall line width of the focused field can be further squeezed by the diffracted field, which is focused at a time half of the period earlier than the focused field, connecting with the transmitting field, which is focused at a time half of the period later, as shown in Fig. 2(c-d). In order to form this field connection, there is a requirement on the width of the central metal strip and the field diffraction has to be slowed down by the surface plasma including the effect of the grooves on the structure. The snapshot of the focused $H_z$ field (Fig. 2c) is taken at the moment that there is a peak of focused positive magnetic field outside the surface and near the peak position of the time-averaged Poynting vector in the original $y$ propagation direction shown later in Fig. 4(a). After being focused at $t = t_f$, the light diffracts to forward angles, and the $H_z$ field propagates out while it is still squeezed as shown in Fig. 2(d). Then, the focused light propagates out to the far zone, as also evidenced from the movies of the electric and magnetic fields [29]. This is in a sharp contrast to an evanescent near-field that cannot propagate out and is static, other than its oscillations in time.

The line width of the focused light can be defined by the full-width-half-maximum (FWHM) or the width of the spot of the $H_z$ energy averaged-along-$x$ as twice the position uncertainty that is defined as $\Delta x = (<x^2> - <x>^2)^{1/2}$, where $<f> = \int H_z^2 f \, dx \, / \int H_z^2 \, dx$. Figure 3(a) shows that the FWHM of the time-averaged $H_z$ field energy agrees well with that for the snapshot of $H_z$ field energies. While the peak intensity of the focused light remains to be higher than that of the incident light, the FWHMs at the normalized distance $kr$ up to 4.17 is smaller than the diffraction limit of half the wavelength. The $x$ profiles of the peak focused light at three different times/phases are shown in Fig. 3(b). At the time $t = t_f$, FWHM is 0.286 λ while the width is 0.217 λ when averaged over the focused line; it is 0.320 λ when averaged over the profile. The FWHM becomes 0.394 λ at the time of 0.12 period later and 0.498 λ at the time of $t = t_f$ + 0.29 while the peak of the focused light has reached at a distance of 0.664 λ away from the surface. All the widths discussed above are smaller than the diffraction limit of half the wavelength. Obviously, the diffraction limit has been overcome by a result at the intermediate zone in which there is such a small single-line width occurring with regard to the focused light.

For a far (near) field [30], the electric field is in phase (out of phase) with the magnetic field so that the Poynting vector is not (is) zero. The focused light beyond the diffraction limit of half the wavelength is located at the intermediate zone [30]. To characterize it further, the time-averaged Poynting vector in the original $y$ propagation direction shown on Fig. 4(a) also indicates that the focused fields are propagating and hence are capable of travelling to the far zone [30]. Considering the fact that a travelling light is an indicator of the capability and simplicity of this approach for moving the focal point and the field energy away from the surface, this makes the FAB lens superior to evanescent near-field solutions [5-14] for many critical applications.

The involvement of the near field on the line width of the focused light is quantitatively investigated in-depth. At the middle location of the central metal strip surface, the magnetic field is out of phase with the electric field and in phase with the surface current $-J_x$, as shown in Fig. 4(b). Thus, it is dominated by the near field. But, at the time of $t = t_f + 0.12$, the focused light has moved out. Its magnetic field near the surface is close to zero on average and is a small positive number at the middle. The focused $E_x$ field has a similar contour as the $H_z$ field. The ratio and locations of their peaks determines the impedance Z. Based on the analytical far field relation, the estimated far $E_x$ field is $ZH_z$ that agrees well with the measured $E_x$ as shown in Fig. 4(c). This is one more indication [30] for the focused light to be dominated by the radiative field, as also evidenced by the propagation of the $E_x$ field shown in Fig. 4(d). At the time 0.01 later, the magnetic field at the middle of the central metal strip has dropped to a negative value very close to zero. As shown in Fig. 4(d), the estimated far $E_x$ field along $kr$ based on the analytical theory is in good agreement with the focused field obtained from the simulation. The near $E_x$ field is yielded from their difference and is decreasing away from the surface as expected with the length of half the field energy being $kr = 0.476$ or less than 0.1 $\lambda$; that is consistent with the length scale of the near zone measured by NSOM [31-32]. The near field is smaller than the far field at $kr > 1$. The effect of the near field is negligible at the intermediate zone of $2 < kr < 4$, where the line width of the focused light is smaller than that of the diffraction limit.

Beyond the diffraction limit, the focusing of light is intellectually intriguing and important for application possibilities. It may be employed to manipulate and image biomoleculars at a higher precision, resolution and depth with propagating light, to sense the structure and dynamics of biological and physical systems at a smaller scale, to diagnose and modify material surfaces with greater precision, to remove the limit on photolithography, which is the key issue preventing the further progress of the semiconductor industry according to Moore's Law, for crafting finer circuits, to produce and read smaller spots for optical storage, to squeeze light for optical detection and into photonic and plasmonic circuits [33], to connect optical systems and finer electronic circuits, among many others. The physical mechanisms might be used in applications required for the processing of optical information and thus in communication and optical computing processes.

In summary, the physical mechanisms of the innovative approach using a miniature FAB lens are demonstrated to focus light to a single-line with its width beyond the lower limit of diffraction in the intermediate zone of $2 < kr < 4$. It is quantitatively verified that the involvement of near-field in the focused fields is negligible. As of result of being able to propagate light, this scheme is superior to near-field proposals. Besides the academic interest generated by the physical mechanisms and the approach, the light focusing process is expected to open up a wide range of application possibilities, especially with regard to the capabilities of the focused light being able to propagate, moving the focal point away from the surface and reducing the sizes of the focused light and the devices corresponding.

**Acknowledgements** This work was supported by NCKU Project of Promoting Academic Excellence & Developing World Class Research Centers.

*chenkr@mail.ncku.edu.tw

## Figure Captions

Fig. 1  (Color online) The schematic structure of the aperture and the approach. (a) Schematic diagram of the approach, aperture structure and the paths of the light transmitted, bent and focused, including the generator of sub-limit wave function at the central area. (b) The schematic structure of the aperture on a silver film used in the FDTD simulation. The depth, the width and the distance in between of the periodical grooves at the incident side are 80 nm, 200 nm and 200 nm, respectively; the slit width is 80 nm; both the width and depth of grooves at the exit side is 80 nm; the distance between the slit and the exit groove is 160 nm; the film thickness is 280 nm; the thickness and the width of the central film are 200 nm and 320 nm, respectively. In order to illuminate the focus beyond the limit, the central metal width of 320 nm and the exit width of 480 nm are larger than the diffraction limit of half the wavelength, 316.5 nm.

Fig. 2  (Color online) The contour plots of the fields. (a) The time-averaged contours of the $E_x$ energy. (b) The time-averaged contours of the Poynting vector in the $x$ direction. (c) The snapshot of the focused $H_z$ field at $t = t_f$. (d) The $H_z$ field at $t = t_f + 0.12$. All are normalized to the incident light.

Fig. 3  (Color online) The profiles and the width of the focused light. (a) The FWHM vs. normalized $r$ profiles of the snapshot of the $H_z$ field energy (blue squares) and the time-averaged $H_z$ field energy (red curve), where $r$ is the $y$ distance from the metal surface. (b) The profiles of the focused $H_z$ field energy at $t = t_f$ (red curve), $t_f + 0.12$ (blue dashes) and $t_f + 0.29$ (green dots).

Fig. 4  (Color online) The Poynting vector contours and the field profiles. (a) The time-averaged contours of the Poynting vector in the $y$ direction. (b) The temporal profiles of the magnetic field $H_z$ (red curve), the electric field $E_x$ (blue dashes) and the current $J_x$ (green dots) at $x = 0$ of the central metal surface. (c) The $x$ profiles of peak $E_x$ (blue dashes) and $ZH_z$ (red curve) fields at $t = t_f + 0.12$; where $|Z| = 0.832$ and the phase is 3 cells or 0.0237 wavelength. (d) The $r$ profiles of the $E_x$ field at $t = t_f$ (light blue short dashes), $t = t_f + 0.13$ (blue dashes), and $t = t_f + 0.29$ (purple dot-dashes) when the FWHM of the $H_z$ energy is still smaller than half the wavelength, as well as the $ZH_z$ field (red curve; the estimated far $E_x$ field; where $|Z| = 0.840$) and the near $E_x$ field (green dots; the difference of the overall and far fields).

Fig. 1

(a)

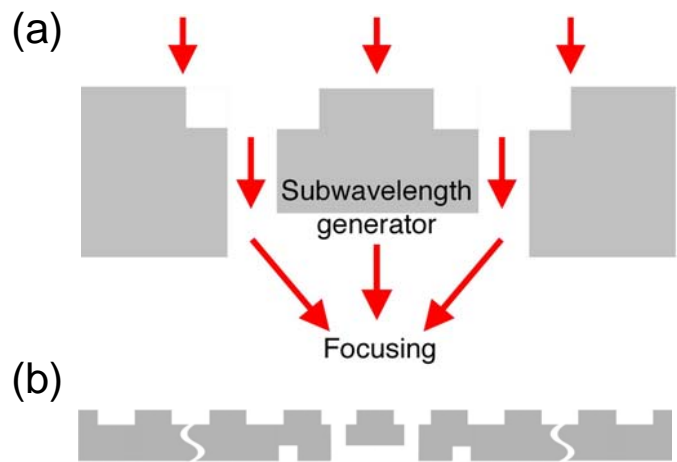

(b)

Fig. 2

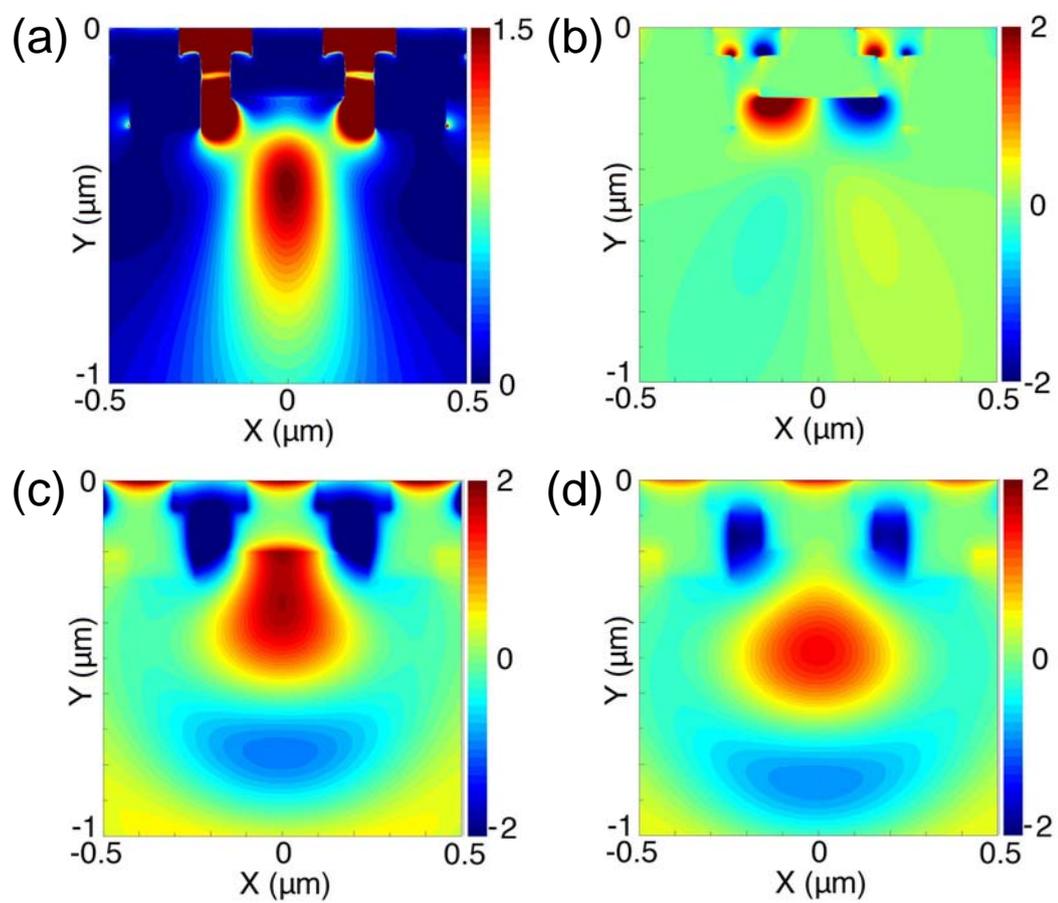

Fig. 3

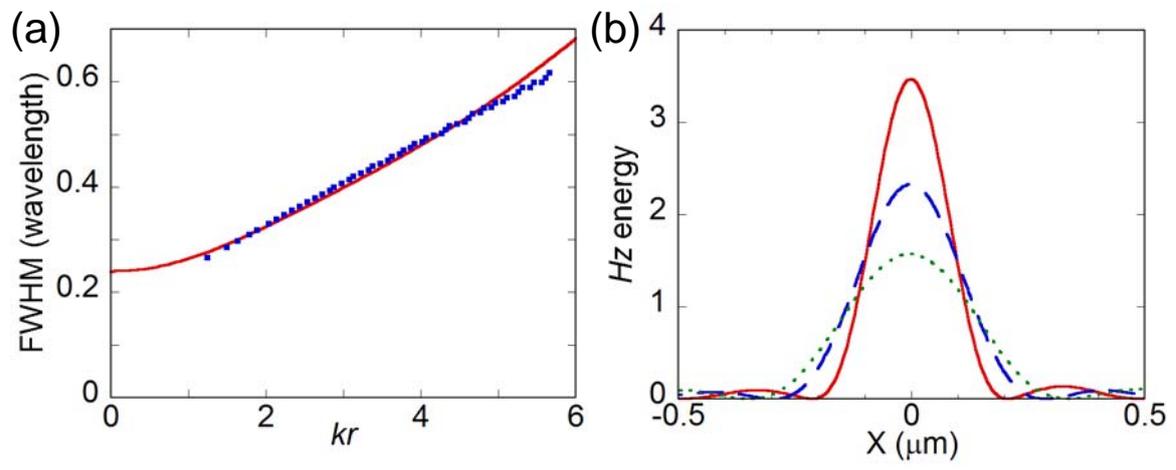

Fig. 4

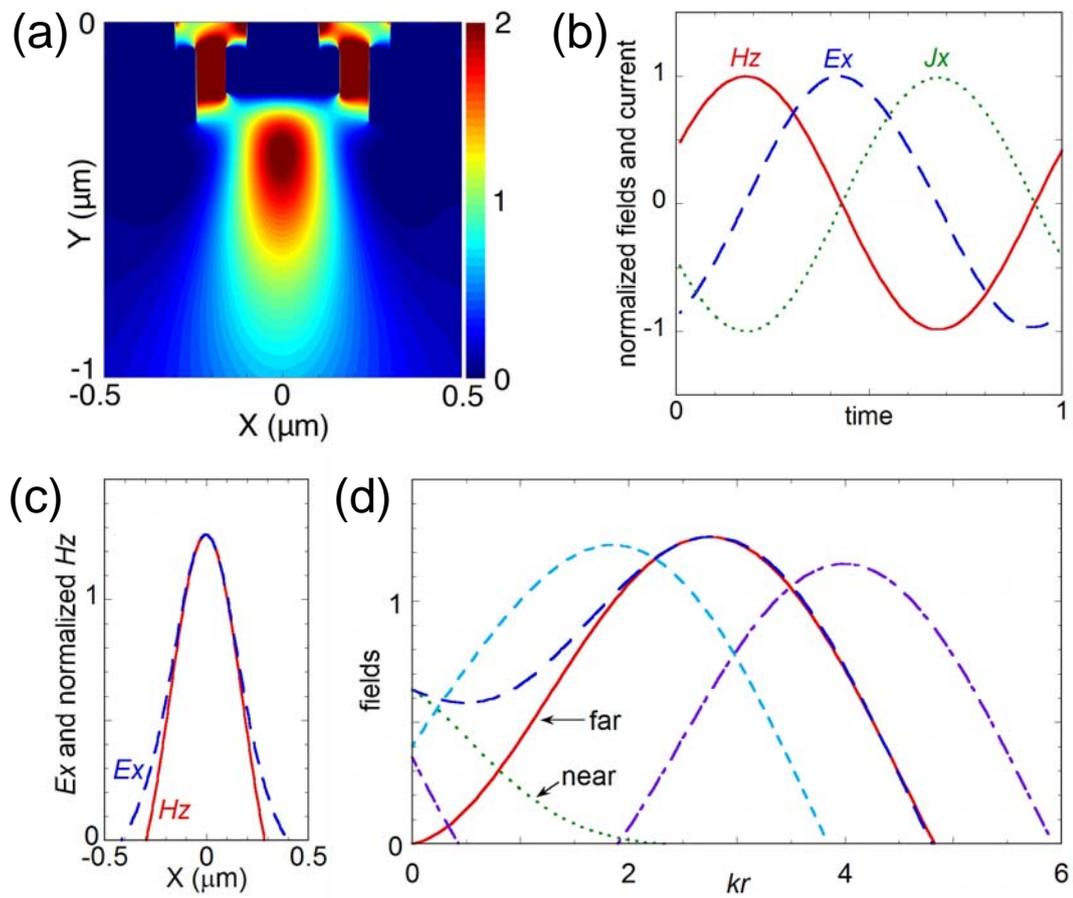